\documentclass[a4paper]{jpconf}
\usepackage{graphicx}
\usepackage{amsmath}
\usepackage{amssymb}
\usepackage{tikz}
\usepackage{siunitx}
\begin{document}
\title{Analytical energy spectra and wake effects for relativistic dielectric laser accelerators}

\author{Thilo Egenolf and Uwe Niedermayer}

\address{Institut f\"ur Teilchenbeschleunigung und Elektromagnetische Felder (TEMF),\\Technische Universit\"at Darmstadt, Schlossgartenstrasse 8, D-64289 Darmstadt, Germany}

\ead{egenolf@temf.tu-darmstadt.de}

\begin{abstract}
Dielectric laser acceleration (DLA) is one of the advanced concepts for more compact accelerators. DLA gratings have apertures and period lengths within the range of optical wavelengths. Phase stability and wakefield effects are thus crucial for upcoming experiments with relativistic electrons. For this, we present a method to analytically calculate energy spectra for comparison with measurements in order to reconstruct the phase of injection into the DLA grating structure. Knowing the injection phase is important for both alignment and interpretation of measured data. Furthermore, we estimate the effects of wakefields on bunches which are coherently accelerated in a DLA. We are calculating the energy spectrum affected by the longitudinal wake by evaluating the analytical description numerically and give estimates for the transverse kicks of an off-centered injected electron bunch.
\end{abstract}

\section{Introduction}
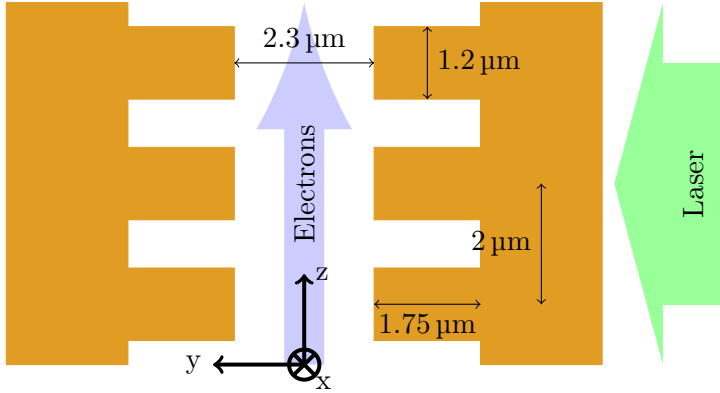
\begin{figure}[htp]
\centering
\begin{tikzpicture}[scale=0.8,rotate=90]
\definecolor{mathematicablue}{rgb}{0.3684,0.5068,0.7098};
\definecolor{mathematicayellow}{rgb}{0.8807,0.611,0.1421};

\def\periodlength{2};
\def\toothwidth{1.2};
\def\toothheight{1.75};
\def\gap{2.3};
\def\basis{2};

\draw[mathematicayellow, fill=mathematicayellow] (-0.5*\periodlength,-\gap/2-\toothheight-\basis) rectangle (2.5*\periodlength,-\gap/2-\toothheight);
\draw[mathematicayellow, fill=mathematicayellow] (-0.5*\periodlength,\gap/2+\toothheight) rectangle (2.5*\periodlength,\gap/2+\toothheight+\basis);
\foreach \x in {0,...,2}
    {\draw[mathematicayellow, fill=mathematicayellow] (\x*\periodlength-\toothwidth/2,-\gap/2-\toothheight) rectangle (\x*\periodlength+\toothwidth/2,-\gap/2);
    \draw[mathematicayellow, fill=mathematicayellow] (\x*\periodlength-\toothwidth/2,\gap/2) rectangle (\x*\periodlength+\toothwidth/2,\gap/2+\toothheight);}
    
\draw[->, >=latex, blue!20!white, line width=15pt]   (-\periodlength/2,0) to node[black,rotate=90]{Electrons} (2.5*\periodlength,0) ;

\path[fill=green!40!white] (\periodlength,-\gap/2-\toothheight-1.1*\basis) -- (-0.5*\periodlength,-\gap/2-\toothheight-1.5*\basis) -- (2.5*\periodlength,-\gap/2-\toothheight-1.5*\basis) -- cycle;
\path[fill=green!40!white] (0,-\gap/2-\toothheight-1.5*\basis) rectangle (2*\periodlength,-\gap/2-\toothheight-2*\basis) node[black,rotate=90,pos=0.5]{Laser};
%\draw[->, >=latex, green!20!white, line width=50pt]   (\periodlength,-\gap/2-\toothheight-2*\basis) to node[black,rotate=90]{Laser} (\periodlength,-\gap/2-\toothheight-\basis) ;

\draw[black, <->] (2*\periodlength,-\gap/2) -- (2*\periodlength,\gap/2) node[pos=0.5,above]{\SI{2.3}{\micro\metre}};
\draw[black, <->] (0,-\gap/2-\toothheight-\basis/2) -- (\periodlength,-\gap/2-\toothheight-\basis/2) node[pos=0.5,left]{\SI{2}{\micro\metre}};
\draw[black, <->] (2*\periodlength-\toothwidth/2,-\gap/2-\toothheight/2) -- (2*\periodlength+\toothwidth/2,-\gap/2-\toothheight/2) node[pos=0.5,right]{\SI{1.2}{\micro\metre}};
\draw[black, <->] (0,-\gap/2-\toothheight) -- (0,-\gap/2) node[pos=0.5,below]{\SI{1.75}{\micro\metre}};

\draw[black, ->, line width=1.5pt] (-0.5*\periodlength,0) -- (0.25*\periodlength,0) node[right]{z};
\draw[black, ->, line width=1.5pt] (-0.5*\periodlength,0) -- (-0.5*\periodlength,0.75*\periodlength) node[left]{y};
\draw[black, line width=1.5pt] (-0.5*\periodlength,0) circle (0.25) node[below right]{x};
\draw[black, line width=1.5pt] (-0.5*\periodlength+0.176,0.176) -- (-0.5*\periodlength-0.176,-0.176);
\draw[black, line width=1.5pt] (-0.5*\periodlength-0.176,0.176) -- (-0.5*\periodlength+0.176,-0.176);

\end{tikzpicture}\hspace{2pc}
\begin{minipage}[b]{12pc}
\caption{\label{fig:RecDimensions}Sketch of a dual-layer rectangular grating structure. The structure is made of fused silica (${\varepsilon_r=2.13}$).}
\end{minipage}
\end{figure}

In the concept of dielectric laser acceleration (DLA)~\cite{England2014DielectricAccelerators} electrons are accelerated in dielectric nanostructures which are powered by ultra-short laser pulses in the optical regime. Experiments with relativistic electrons are planned or have already been carried out at various international research facilities within the "Accelerator on a chip" (ACHIP) collaboration~\cite{Mayet2018SimulationsSINBAD,Kuropka2019PlansSINBAD,Kuropka2018FullSINBAD,Prat2017OutlineSwissFEL,Cesar2018High-fieldAccelerator,Cesar2018EnhancedLaser,Peralta2013DemonstrationMicrostructure,Wootton2016DemonstrationPulses}. The experiments use an externally accelerated electron beam from a conventional photoinjector or radio-frequency (RF) linac to inject in a DLA grating structure. Two of these planned experiments aim for coherent acceleration of an injected electron bunch. At SwissFEL (PSI) the RF linac is used to accelerate electron bunches to \SI{3}{\giga\electronvolt} and compress them~\cite{Prat2017OutlineSwissFEL}. The proposed experiment at DESY's SINBAD facility applies the ARES linac to accelerate an electron bunch to \SI{52}{\mega\electronvolt} and transform the bunch into a train of ultra-short microbunches with \SI{350}{\atto\second} FWHM bunch length by laser modulation~\cite{Mayet2017ASINBAD}. The goal of both experiments is to coherently accelerate the microbunches in a dual-layer rectangular grating with \SI{2}{\micro\metre} period length. A sketch of such a structure made of fused silica is shown in Fig.~\ref{fig:RecDimensions}. Since at SINBAD both the modulator and the DLA grating are driven by the same laser and at SwissFEL the drive laser is synchronized with the RF, phase synchronization between the microbunches and the accelerating field can be established. However, the phase of injection into the DLA grating is not known or measurable in advance. To get around this, we present an analytical scheme to reconstruct the injection phase from the measured energy spectrum after the DLA interaction in Sec.~\ref{sec:energySpectrum}. Furthermore, the proposed microbunch charge of a few femtocoulomb leads to a significant wake field in the DLA grating. The effects of this wake are estimated in Sec.~\ref{sec:wake}.

\section{\label{sec:energySpectrum}Injection phase reconstruction from energy spectrum}
\begin{figure}[h]
\begin{minipage}[t]{17.8pc}
\includegraphics[width=17.8pc]{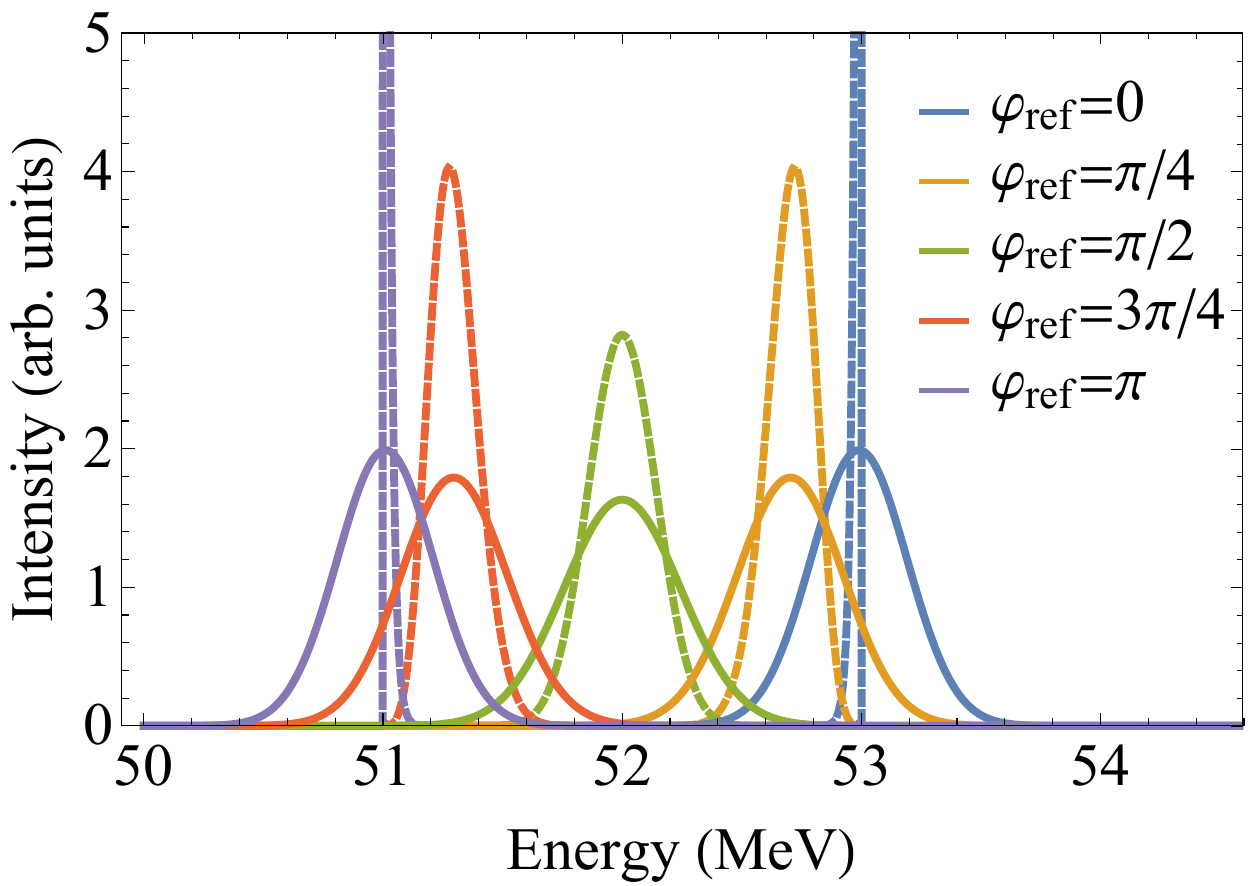}
\caption{\label{fig:energySpectrumPhase}Energy spectra after DLA interaction for different injection phases. The dashed lines show the energy spectra without initial energy spread, the solid lines show the energy spectra with \SI{200}{\kilo\electronvolt} rms energy spread.}
\end{minipage}\hspace{2pc}%
\begin{minipage}[t]{17.8pc}
\includegraphics[width=17.8pc]{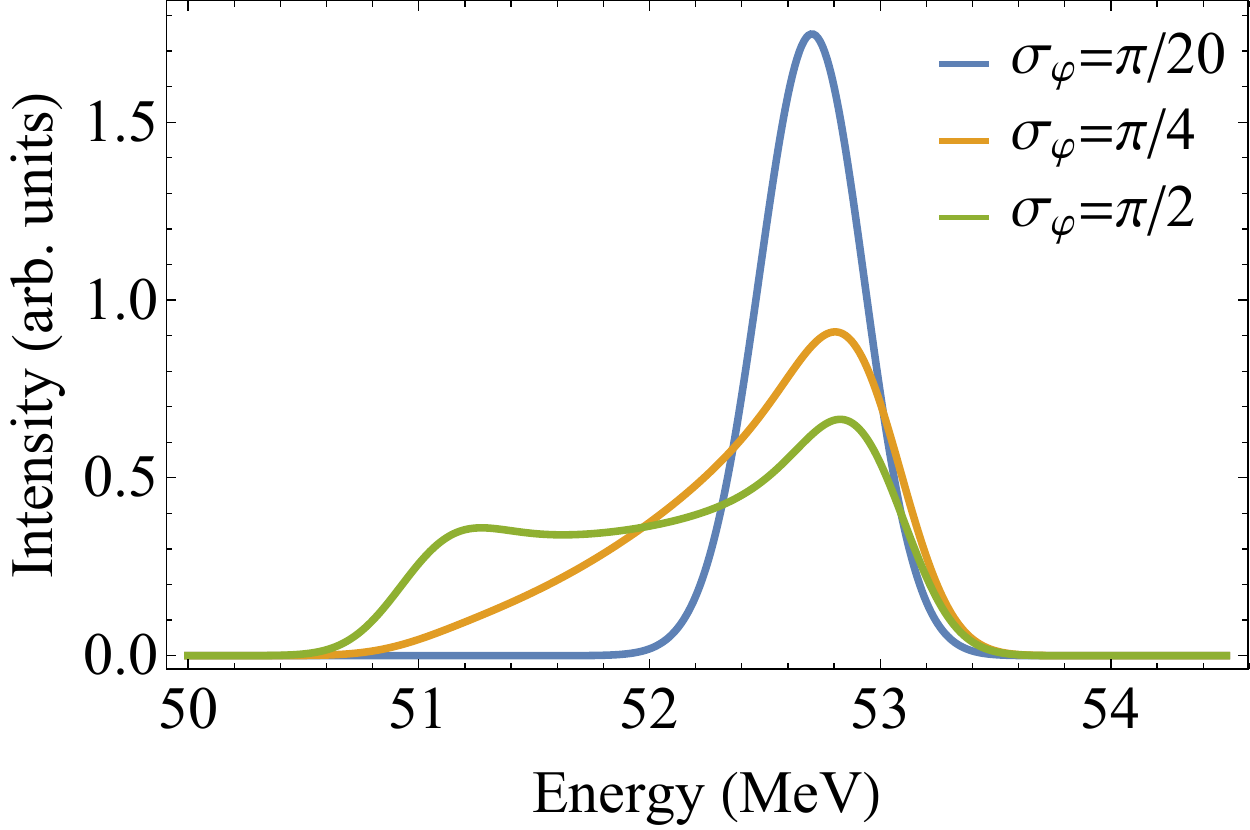}
\caption{\label{fig:energySpectrumWidth}Energy spectra after DLA interaction for bunches with different width injected at $\varphi=\pi/4$.\\\\\\}
\end{minipage} 
\end{figure}

In order to reconstruct the injection phase from a measured energy spectrum, knowledge of the initial bunch distribution and the DLA interaction is required. We will construct the energy spectrum for these parameters, such that parametric fitting of the analytical formula to a measured energy spectrum allows to retrieve the injection phase. Suppose the bunch is initially Gaussian distributed, i.e. the uncorrelated two-dimensional probability density function of the initial bunch is given by
\begin{equation}
    \lambda_{0}\left(\varphi,w\right)=\frac{1}{\sqrt{2\pi}\sigma_\varphi}\frac{1}{\sqrt{2\pi}\sigma_w}\exp\left(-\frac{\left(\varphi-\varphi_\textrm{ref}\right)^2}{2\sigma_\varphi^2}\right)\exp\left(-\frac{\left(w-w_\textrm{ref}\right)^2}{2\sigma_w^2}\right)
    \label{eq:distribution}
\end{equation}
with standard deviations $\sigma_\varphi$ and $\sigma_w$ and means $\varphi_\textrm{ref}$ and $w_\textrm{ref}$, where $\varphi$ describes the injection phase and $w$ the energy of each particle. The energy of each particle after an interaction length~l is given by the function $W:\mathbb{R}^2\rightarrow\mathbb{R}$ as
\begin{equation}
     W\left(\varphi,w\right)=qle_1\cos\left(\varphi\right)+w
     \label{eq:energy}
\end{equation}
with the particle charge $q$ and the peak acceleration gradient $e_1$. For this, we have neglected a change in phase within the interaction and also a transverse dependence of the acceleration gradient. These assumptions are fulfilled for a relativistic electron bunch, where
%and an interaction length of a few hundred DLA periods. 
the acceleration gradient is proportional to $\cosh\left(2\pi y/\left(\lambda_L\beta\gamma\right)\right)$ with the laser wavelength $\lambda_L$
%which results in an almost constant gradient over the entire gap
~\cite{Niedermayer2017BeamScheme}. 
%This confirms the second assumption.
In order to map the initial bunch distribution to the energy distribution after the interaction length, we have to use probability theory and solve the integral
\begin{equation}
    \lambda_{l}\left(\tilde{w}\right)=\int_{-\infty}^\infty\frac{\lambda_{0}\left( \varphi\left(\tilde{w},w\right),w\right)}{\left|\frac{\partial W\left(\varphi,w\right)}{\partial\varphi}\right|}dw,
    \label{eq:distributionW}
\end{equation}
where $\tilde{w}=W\left(\varphi,w\right)$. To solve this, we need the  inverse function $\varphi=\arccos{\left[\left(\tilde{w}-w\right)/\left(qle_1\right)\right]}$ and the derivative of Eq.~\ref{eq:energy}, $\frac{\partial W\left(\varphi,w\right)}{\partial\varphi}=-qle_1\sin\left(\varphi\right)$. This allows us to calculate Eq.~\ref{eq:distributionW} as
\begin{equation}
    \lambda_{l}\left(\tilde{w}\right)=\frac{1}{\sqrt{2\pi}\sigma_w}\int_{-\infty}^\infty\Lambda\left(\tilde{w}\right)\exp\left(-\frac{\left(w-w_\textrm{ref}\right)^2}{2\sigma_w^2}\right)dw
    \label{eq:distributionW2}
\end{equation}
with
\begin{equation}
    \Lambda\left(\tilde{w}\right)=\frac{1}{\sqrt{2\pi}\sigma_\varphi}\frac{1}{\sqrt{\left(qle_1\right)^2-\tilde{w}^2}}\exp\left(-\frac{\left(\arccos{\left(\frac{\tilde{w}}{qle_1}\right)}-\varphi_\textrm{ref}\right)^2}{2\sigma_\varphi^2}\right).
    \label{eq:deltadistribution}
\end{equation}
Note that Eq.~\ref{eq:deltadistribution} is the mapped probability function of a bunch which initial energy distribution is a Dirac delta function, $\Lambda\left(\tilde{w}\right)= \left.\lambda_l\left(\tilde{w}\right)\right|_{\sigma_w\rightarrow0}$, and therefore Eq.~\ref{eq:distributionW2} is the convolution of this mapped probability function with the Gaussian energy distribution. Furthermore, the energy distribution after interaction for an arbitrary initial energy distribution is also given by the convolution of $\Lambda\left(\tilde{w}\right)$ with the initial energy distribution. Figure~\ref{fig:energySpectrumPhase} shows examples of energy spectra of \SI{45}{\nano\metre} bunches after a DLA interaction with different injection phases. The energy spectra of bunches with different width injected at the same phase are shown in Fig.~\ref{fig:energySpectrumWidth}. The plot shows coherent acceleration for the shortest bunch and the characteristic double horn structure for the longest bunch.
For a well-known initial longitudinal phase space distribution, either by measurements or by simulations, the analytical formulas can be used to calculate the energy spectrum after interaction for different injection phases and determine the correct phase by comparison with a measured energy spectrum.
\begin{figure}[h]
\includegraphics[width=17.8pc]{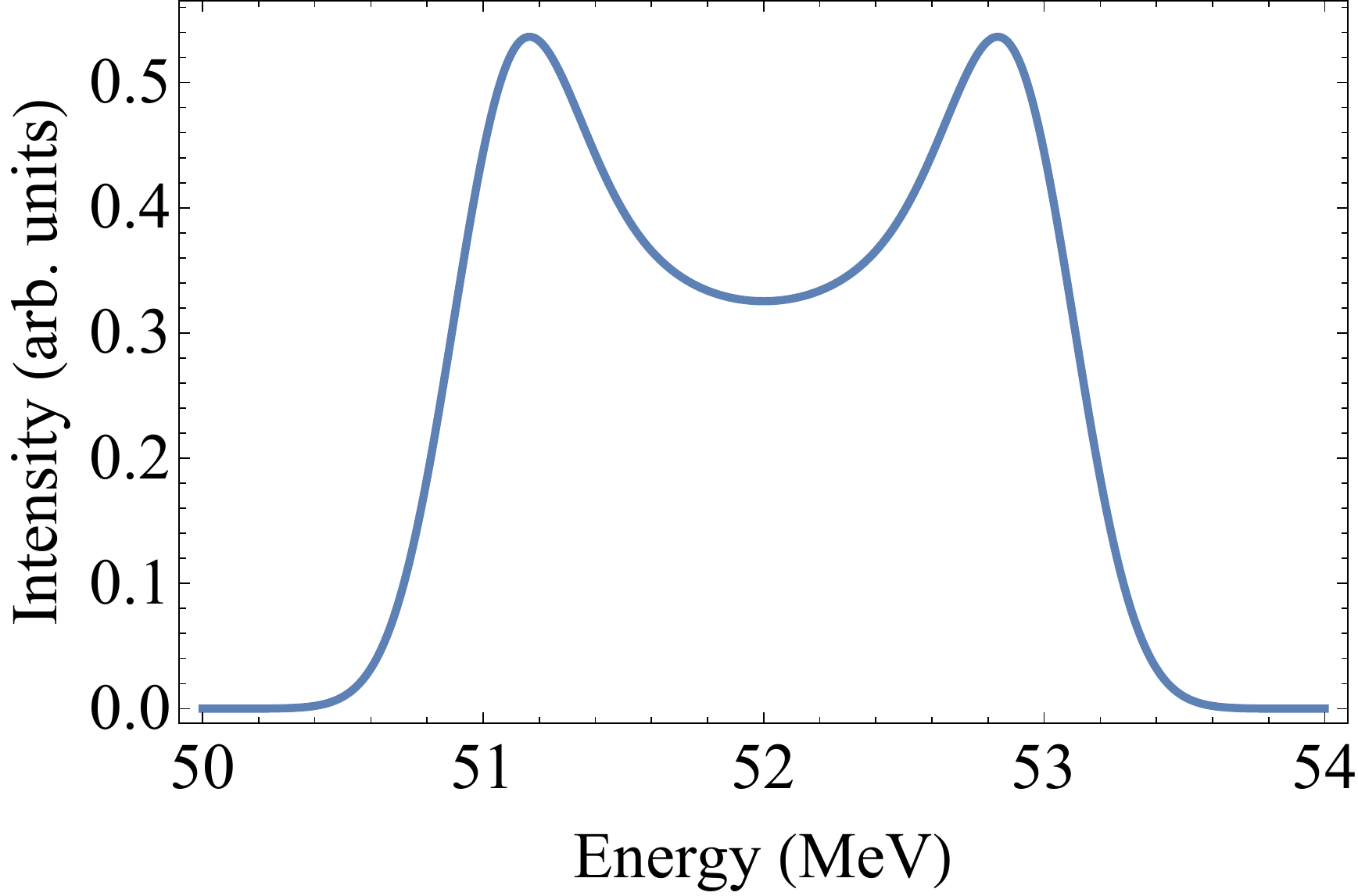}
\hspace{2pc}
\begin{minipage}[b]{17.8pc}
\caption{\label{fig:energyspectrumLongBunch}Double horn energy spectrum after DLA interaction for a bunch much longer than the grating period\\}
\end{minipage}
\end{figure}
Double horn energy spectra have already been measured in DLA experiments with bunches much longer than the period length of the used DLA grating~\cite{Peralta2013DemonstrationMicrostructure,Wootton2016DemonstrationPulses}. The energy spectra of such experiments can also be described analytically by the presented formulas (cf. Fig.~\ref{fig:energyspectrumLongBunch}). Furthermore, it is also possible to solve Eq.~\ref{eq:distributionW} numerically if the longitudinal bunch distribution cannot be described analytically. Double-horn energy spectra from sub-relativistic DLA experiments~\cite{Leedle2015DielectricStructures,Yousefi2019DielectricReflector} can also be reconstructed, provided the interaction length is sufficiently short and the transverse particle distribution and the transverse dependence of the gradient is also being taken into account by another convolution.

\section{\label{sec:wake}Wakefield effects}
\begin{figure}[h]
\begin{minipage}{17.8pc}
\includegraphics[width=17.8pc]{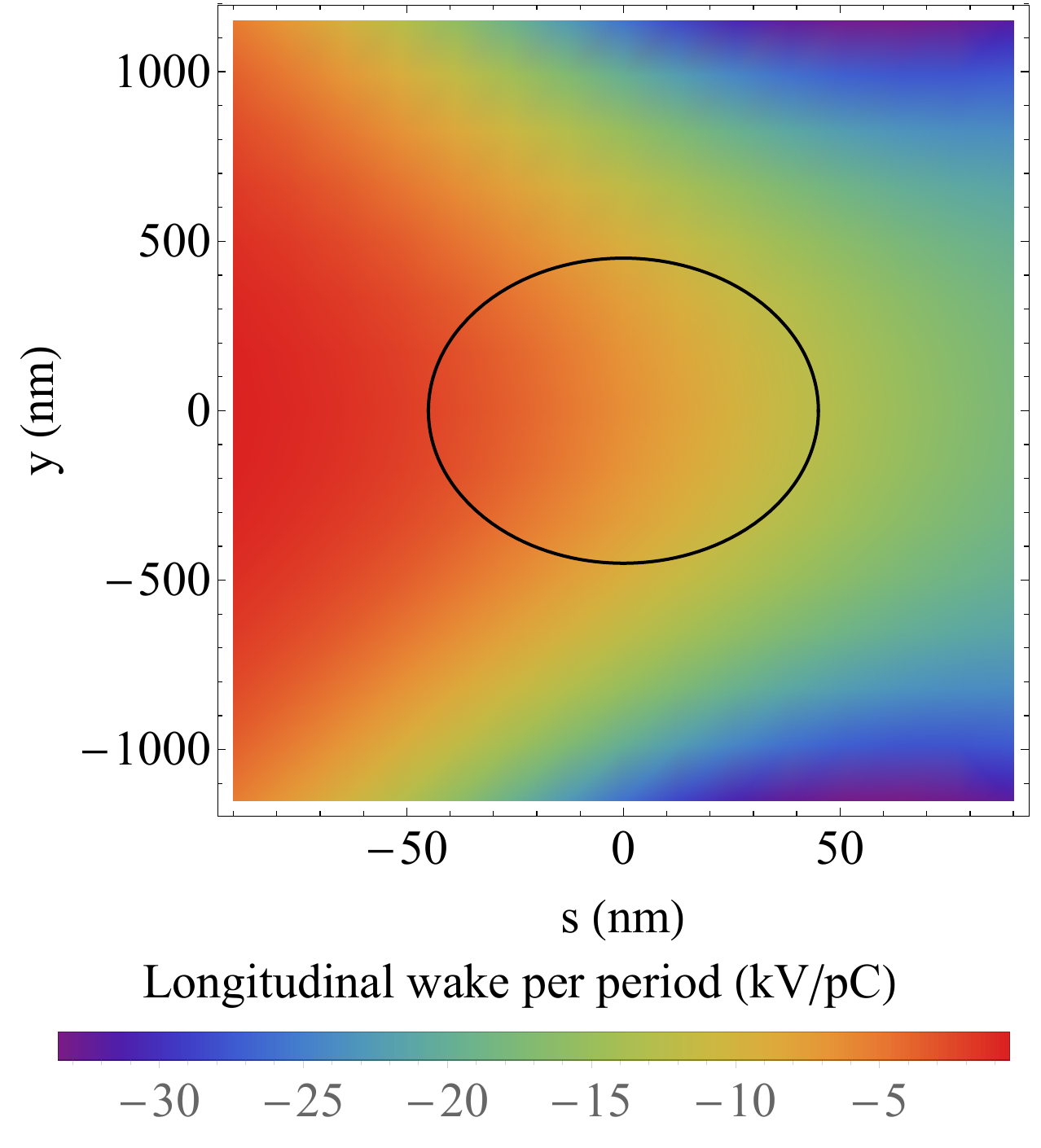}
%\caption{\label{fig:longWake}Longitudinal wake potential at $x=0$ per DLA period (\SI{2}{\micro\metre}) of a Gaussian bunch with $\sigma_s=\SI{45}{\nano\metre}$ and $\sigma_{x,y}=\SI{450}{\nano\metre}$ given by the black ellipse in a dual-layer rectangular grating structure}
\end{minipage}\hspace{2pc}%
\begin{minipage}{17.8pc}
\includegraphics[width=17.8pc]{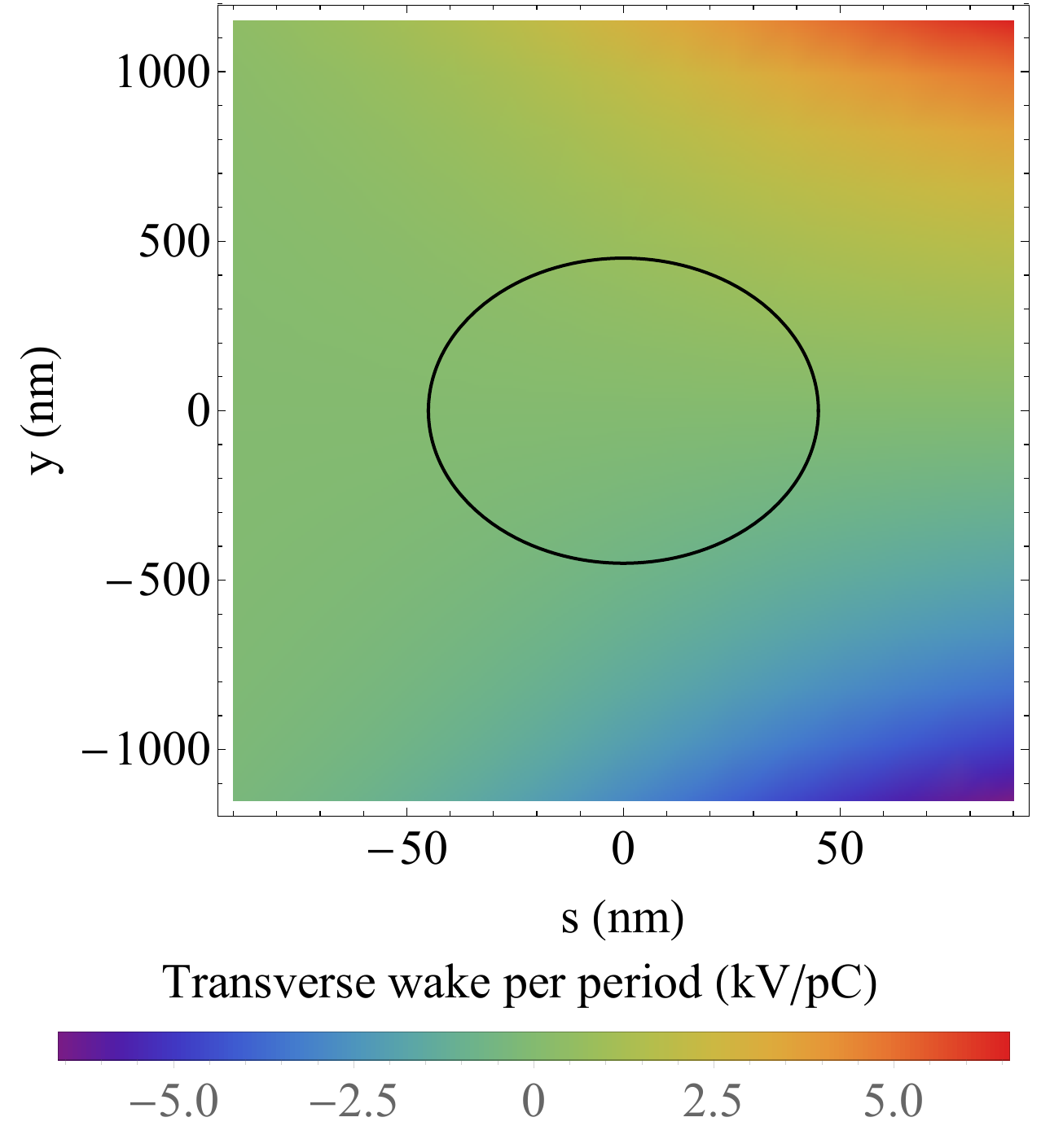}
\end{minipage}
\caption{\label{fig:transWake}Longitudinal and transverse wake potentials at ${x=0}$ per DLA period (\SI{2}{\micro\metre}) of a Gaussian bunch with $\sigma_s=\SI{45}{\nano\metre}$ and $\sigma_{x,y}=\SI{450}{\nano\metre}$ given by the black ellipse in a dual-layer rectangular grating structure}
%\end{minipage} 
\end{figure}

\begin{figure}[h]
\begin{minipage}[t]{17.8pc}
\includegraphics[width=17.8pc]{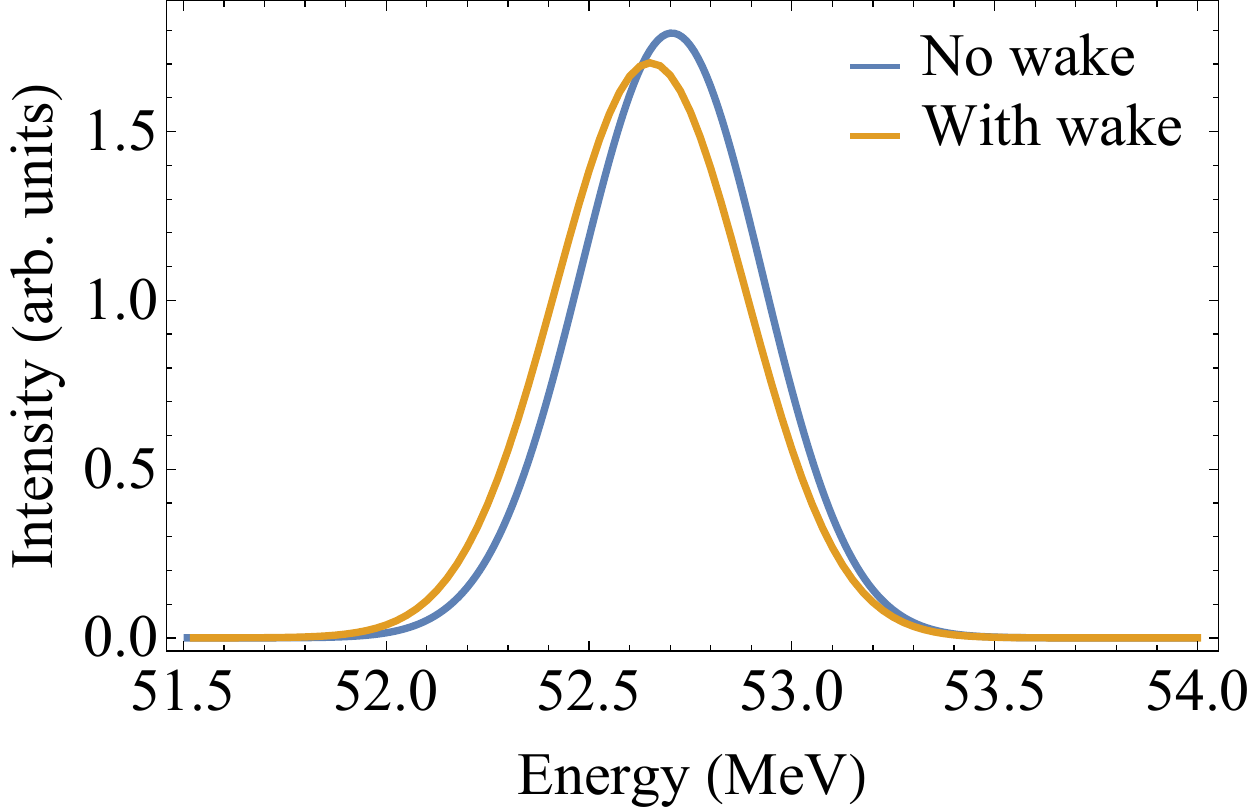}
\caption{\label{fig:energySpectrumWake}Energy spectrum after 500 DLA periods with and without longitudinal wakefield. The peak acceleration gradient is $e_1=\SI{1}{\giga\electronvolt\per\metre}$ and the injection phase is $\phi_\textrm{ref}=\pi/4$.}
\end{minipage}
\hspace{2pc}%
\begin{minipage}[t]{17.8pc}
\includegraphics[width=17.8pc]{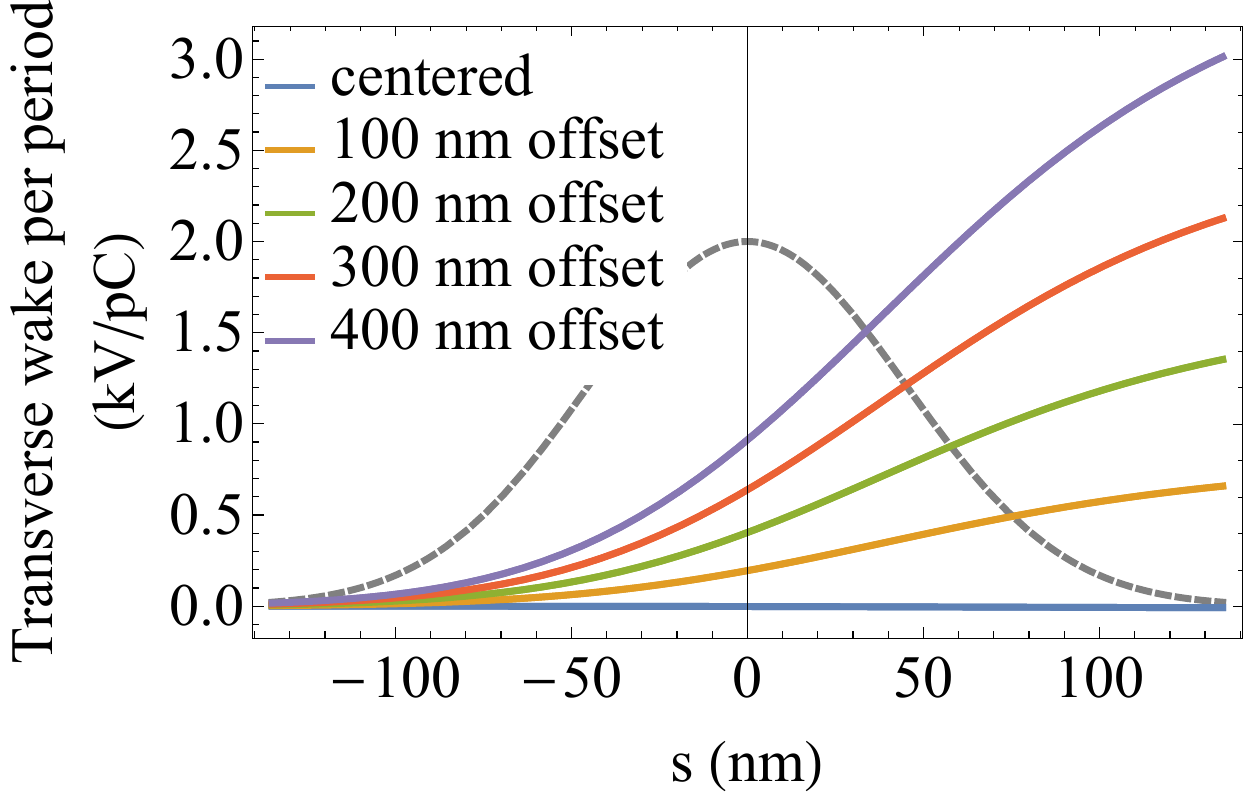}
\caption{\label{fig:transWakeOffset}Transverse wake potential per DLA period (\SI{2}{\micro\metre}) of a Gaussian bunch with injection offset $y_0$ at $x=0$ and $y=y_0$. The bunch distribution is shown in dashed gray.}
\end{minipage}
\end{figure}

%Additionally to the laser field, which could be described analytically, the wakefield as self-field 
The wake field created by the electron beam in the DLA structure can decelerate, defocus, and deflect the beam depending on bunch charge and distribution. To estimate these effects, the fields are simulated by the dedicated wakefield solver of CST Studio Suite~\cite{DassaultSystemesDeutschlandGmbH2019CST2019} using the rigid beam assumption. Postprocessing steps to calculate the three-dimensional wake fields for arbitrary bunch distributions are detailed in~\cite{Egenolf2019IntensityStructures}. Assuming that microbunches, as e.g. created by the SINBAD microbunching scheme, are Gaussian distributed and injected in the center of the channel, the corresponding wakefields are plotted in the Fig.~\ref{fig:transWake}.
The longitudinal wake acts as decelerating force. A wake per DLA period of \SI{-10}{\kilo\volt\per\pico\coulomb}, for example, corresponds to a gradient of \SI{-5}{\giga\electronvolt\per\pico\coulomb} or an energy loss of \SI{5}{\mega\electronvolt\per\pico\coulomb} in an interaction length of \SI{1}{\milli\metre}. Figure~\ref{fig:energySpectrumWake} shows the energy spectrum after the DLA interaction, with and without the effect of the longitudinal wake. 
Without wake, the spectrum was obtained by evaluating Eq.~\ref{eq:distributionW} analytically for a Gaussian bunch distribution with a bunch length of \SI{45}{\nano\metre} and \SI{200}{\kilo\electronvolt} energy spread. 
The wake was included by adding the corresponding energy loss (evaluated at the center of the channel) to the energy gain in Eq.~\ref{eq:energy}. This leads to a slight shift of the spectrum to smaller energy. The transverse wake in $y$-direction acts both defocusing in general and coherently deflecting for bunches with off-center injection. Figure~\ref{fig:transWakeOffset} shows the transverse wake along the bunch center ($x=0$,$y=y_0$) for various injection offsets $y_0$. 
Evaluating the transverse wake at the bunch center ($s=0$) normalized to the offset yields $W'_y=\SI{1.05}{\peta\volt\per\metre\squared\per\pico\coulomb}$ which results in the kick
\begin{equation}
     \Delta y'=\frac{e\, q_\mathrm{bunch}}{p_{z0}\beta c_0}W'_yy_0l\approx\SI{20e3}{\per\metre\per\pico\coulomb}q_\mathrm{bunch}y_0,
\end{equation}
where $e$ is the elementary charge and $p_{z0}$ is the reference momentum.
The absence of longitudinal and transverse motion in a stiff relativistic beam allows to lump all kicks in a millimeter-length DLA structure to be lumped in a single kick after the DLA interaction.
Calculating this kick for specific parameters determines whether the wake of the DLA has a measurable or deteriorating effect. 
%is measurable or affects the beam dynamics after the DLA interaction.

\section{Conclusion}
The planned experiments at SwissFEL and at the SINBAD facility will enable coherent acceleration of highly relativistic electrons in a dielectric laser acceleration grating for the first time. In order to determine injection phase of bunches significantly shorter than the DLA period, we presented a scheme to calculate the downstream energy spectrum  analytically. These spectra can be parametrically fitted to measured ones and thus the injection phase can be retrieved. In addition to a Gaussian bunch distribution and a cosine acceleration potential, the equations can also be solved numerically using an arbitrary bunch distribution and an invertible energy gain function, such as the numerically obtained longitudinal wake potential. For the presented example, the longitudinal wakefields are small compared to the proposed acceleration gradients. Also the transverse wake does not lead to a significant deflection within the interaction length of \SI{1}{\milli\metre}. However, the energy loss and the transverse kicks are possibly measurable. Moreover, for a longer interaction length they cannot be neglected in any case. Although the wakefields show a strong transverse dependence, we only considered the wake in the bunch center for a first estimate. Tracking of the particle distribution in the wakefields including the transverse dependence and analysis of coherent beam instabilities will be presented in another paper in the near future~\cite{Egenolf2019WakefieldStructures}.

\ack
The authors would like to thank the ACHIP collaborators from DESY and PSI for inspiring discussions and sharing their parameters. This work is funded by the Gordon and Betty Moore Foundation (Grant No. GBMF4744) and the German Federal Ministry of Education and Research (Grant No. FKZ:05K16RDB).

\section*{References}
\bibliographystyle{iopart-num}
\bibliography{referencesEAAC.bib}

\providecommand{\newblock}{}
\begin{thebibliography}{10}
\expandafter\ifx\csname url\endcsname\relax
  \def\url#1{{\tt #1}}\fi
\expandafter\ifx\csname urlprefix\endcsname\relax\def\urlprefix{URL }\fi
\providecommand{\eprint}[2][]{\url{#2}}

\bibitem{England2014DielectricAccelerators}
England R~J {\em et~al.\/} 2014 {\em Rev. Mod. Phys.\/} {\bf 86} 1337--1389

\bibitem{Mayet2018SimulationsSINBAD}
Mayet F, Assmann R, B{\"{o}}dewadt J, Brinkmann R, Dorda U, Kuropka W, Lechner
  C, Marchetti B and Zhu J 2018 {\em Nucl. Instrum. Meth. A\/} {\bf 909}
  213--216

\bibitem{Kuropka2019PlansSINBAD}
Kuropka W, A{\ss}mann R, Burkart F, Cankaya H, Dorda U, Hartl I, K{\"{a}}rtner
  F~X, Lemery F, Marchetti B and Mayet F 2019 {Plans for Dielectric Laser
  Accelerators at SINBAD} {\em 2018 IEEE Advanced Accelerator Concepts
  Workshop, AAC 2018 - Proceedings\/} (Breckenridge, CO, USA)

\bibitem{Kuropka2018FullSINBAD}
Kuropka W, Mayet F, A{\ss}mann R and Dorda U 2018 {\em Nucl. Instrum. Meth.
  A\/} {\bf 909} 193--195

\bibitem{Prat2017OutlineSwissFEL}
Prat E {\em et~al.\/} 2017 {\em Nucl. Instrum. Meth. A\/} {\bf 865} 87--90

\bibitem{Cesar2018High-fieldAccelerator}
Cesar D {\em et~al.\/} 2018 {\em Commun. Phys.\/} {\bf 1} 46

\bibitem{Cesar2018EnhancedLaser}
Cesar D, Maxson J, Musumeci P, Shen X, England R~J, Wootton K~P and Tan S 2018
  {\em Opt. Express\/} {\bf 26} 29216--29224

\bibitem{Peralta2013DemonstrationMicrostructure}
Peralta E~A {\em et~al.\/} 2013 {\em Nature\/} {\bf 503} 91--94

\bibitem{Wootton2016DemonstrationPulses}
Wootton K~P, Wu Z, Cowan B~M, Hanuka A, Makasyuk I~V, Peralta E~A, Soong K,
  Byer R~L and England R~J 2016 {\em Opt. Lett.\/} {\bf 41} 2696--2699

\bibitem{Mayet2017ASINBAD}
Mayet F, Assmann R, Boedewadt J, Brinkmann R, Dorda U, Kuropka W, Lechner C,
  Marchetti B and Zhu J 2017 {A Concept for Phase-Synchronous Acceleration of
  Microbunch Trains in DLA Structures at SINBAD} {\em Proc. of IPAC 2017,\/}
  (Copenhagen, Denmark) pp 3260--3263

\bibitem{Niedermayer2017BeamScheme}
Niedermayer U, Egenolf T and Boine-Frankenheim O 2017 {\em Phys. Rev. Accel.
  Beams\/} {\bf 20} 111302

\bibitem{Leedle2015DielectricStructures}
Leedle K~J, Ceballos A, Deng H, Solgaard O, Pease R~F, Byer R~L and Harris J~S
  2015 {\em Opt. Lett.\/} {\bf 40} 4344--4347

\bibitem{Yousefi2019DielectricReflector}
Yousefi P, Sch{\"{o}}nenberger N, Mcneur J, Koz{\'{a}}k M, Niedermayer U and
  Hommelhoff P 2019 {\em Opt. Lett.\/} {\bf 44} 1520

\bibitem{DassaultSystemesDeutschlandGmbH2019CST2019}
{Dassault Systemes Deutschland GmbH} 2019 {CST Studio Suite 2019}
  \urlprefix\url{http://www.cst.de}

\bibitem{Egenolf2019IntensityStructures}
Egenolf T, Niedermayer U and Boine-Frankenheim O 2019 {Intensity limits by
  wakefields in relativistic dielectric laser acceleration grating structures}
  {\em 2018 IEEE Advanced Accelerator Concepts Workshop, AAC 2018 -
  Proceedings\/} (Breckenridge, CO, USA)

\bibitem{Egenolf2019WakefieldStructures}
Egenolf T, Niedermayer U and Boine-Frankenheim O {Tracking with wakefields in
  dielectric laser acceleration grating structures} {\em submitted to Phys.
  Rev. Accel. Beams\/} (arXiv:1911.03337)

\end{thebibliography}
%\begin{thebibliography}{9}
%\bibitem{iopartnum} IOP Publishing is to grateful Mark A Caprio, Center for Theoretical Physics, Yale University, for permission to include the {\tt iopart-num} \BibTeX package (version 2.0, December 21, 2006) with  this documentation. Updates and new releases of {\tt iopart-num} can be found on \verb"www.ctan.org" (CTAN). 
%\end{thebibliography}

\end{document}